# Evidence for Fast Interlayer Energy Transfer in MoSe$_2$/WS$_2$ Heterostructures

*Daichi Kozawa[a,b,c], Alexandra Carvalho[d,e], Ivan Verzhbitskiy[d,e], Francesco Giustiniano[d,e], Yuhei Miyauchi[a,f], Shinichiro Mouri[a], A. H. Castro Neto[d,e], Kazunari Matsuda[a], Goki Eda[*,d,e,g]*

[a]*Institute of Advanced Energy, Kyoto University, Gokasho, Uji, Kyoto, Japan 611-0011*
[b]*Department of Applied Physics, Waseda University, 3-4-1, Okubo, Shinjuku, Tokyo Japan 169-8555*
[c]*Department of Applied Physics, Nagoya University, Furo-cho, Chikusa-ku, Nagoya Japan 464-8603*
[d]*Department of Physics, National University of Singapore, 2 Science Drive 3, Singapore 117551*
[e]*Centre for Advanced 2D Materials, National University of Singapore, 6 Science Drive 2, Singapore 117546*
[f]*PRESTO, Japan Science and Technology Agency, 4-1-8 Honcho Kawaguchi, Saitama Japan 332-0012*
[g]*Department of Chemistry, National University of Singapore, 3 Science Drive 3, Singapore 117543*

*E−mail: g.eda@nus.edu.sg
*Fax: +65 6777 6126

ABSTRACT

Strongly bound excitons confined in two−dimensional (2D) semiconductors are dipoles with a perfect in−plane orientation. In a vertical stack of semiconducting 2D crystals, such in−plane excitonic dipoles are expected to efficiently couple across van der Waals gap due to strong interlayer Coulomb interaction and exchange their energy. However, previous studies on heterobilayers of group 6 transition metal dichalcogenides (TMDs) found that the exciton decay dynamics is dominated by interlayer charge transfer (CT) processes. Here, we report an experimental observation of fast interlayer energy transfer (ET) in MoSe$_2$/WS$_2$ heterostructures using photoluminescence excitation (PLE) spectroscopy. The temperature dependence of the transfer rates suggests that the ET is Förster−type involving excitons in the WS$_2$ layer resonantly exciting higher−order excitons in the MoSe$_2$ layer. The estimated ET time of the order of 1 ps is among the fastest compared to those reported for other nanostructure hybrid systems such as carbon nanotube bundles. Efficient ET in these systems offers prospects for optical amplification and energy harvesting through intelligent layer engineering.

Energy transfer (ET) is a ubiquitous phenomenon commonly observed in molecular complexes, biological systems, and semiconductor heterostructures[1–6]. In a properly designed heterostructure, energy of an excited state travels across heterointerfaces, resulting in physical responses unique to the system. Intelligent design and choice of materials allow realization of optical and optoelectronic functionalities tailored for specific applications[7]. For example, ET in quantum wells (QW) coupled with quantum dots (QDs) enables efficient electrical pumping of optical excitations in QDs, resulting in enhanced electro–optics[1, 8]. Fluorescence resonance ET in functionalized QDs has also been actively studied for sensing and imaging applications[9].

The dynamics of ET depends on a variety of factors such as overlap between emission and absorption spectrum of donor and acceptor species, their physical separation, temperature, dielectric environment, and relative orientation of donor and acceptor dipole moments[4]. In a van der Waals heterostructure of two–dimensional (2D) semiconductors, Förster–type ET is expected to be efficient due to the parallel orientation[10] and subnanometer separation of in–plane excitons[2, 11] in the neighboring layers. This is a fundamentally unique system in which the 2D dipoles of two distinct materials are at their ultimate proximity and interacting with the highest possible coupling strength. However, commonly studied 2D semiconductors such as $MoS_2$ and $WSe_2$ typically form a type–II heterojunction with a minimal spectral overlap in exciton resonances[12] and the dominant photoinduced relaxation dynamics is believed to be the interlayer charge transfer (CT)[12–16] rather than ET. In this contribution, we show that despite the type–II band alignment and a weak spectral overlap, interlayer ET is the dominant relaxation pathway in $MoSe_2/WS_2$ heterostructures when $WS_2$ layer is excited. We estimate the lower bound of the ET rate to be in picosecond to subpicosecond time scales and show that the process is mediated by strong interlayer dipole–dipole interaction.

Heterobilayers of $MoSe_2/WS_2$ on quartz substrates (Figure 1a) were prepared by dry transfer method (see Methods for details). The two layers are incommensurately stacked as can be seen from the orientation of the crystal edges (Figure 1b). Due to relatively large difference in the lattice constants[17] ($MoSe_2$: 3.313 Å; $WS_2$: 3.197 Å) and random stacking, the two layers are expected to be weakly coupled via van der Waals interaction with minimal perturbation to the electronic band structure of the pristine materials[18, 19]. Our DFT calculations predict that the layers exhibit type–II band alignment with an interlayer band gap that is smaller than that of individual layers (Fig 1c. See also the details in Supporting Note 2, 3 and Figure S4), similar to the case of a $MoS_2/WSe_2$ heterostructure for which the band alignment was experimentally verified[20]. The conduction band electronic states have a minimum in the $WS_2$ layer and valence band electronic states have a maximum in the $MoSe_2$ layer.

Figure 1d shows the differential reflectance spectra ($\Delta R/R$) for $MoSe_2/WS_2$ heterostructure and corresponding isolated monolayers measured at 78 K. These spectra can be interpreted as the absorption spectra[21]. The heterostructure spectrum exhibits six peak features corresponding to the characteristic excitonic series (M–A, M–B, W–A, W–B, etc) of the two materials (M: $MoSe_2$; W: $WS_2$)[19]. It can be seen that the two layers contribute almost equally to the overall absorption. We find that the A and B absorption peaks of the heterostructure are red–shifted by ~20 meV and broadened by 10–20 meV compared to those of isolated monolayer $MoSe_2$ and $WS_2$.

Photoluminescence (PL) spectrum of the heterobilayer (Figure 1e) consists of two predominant peaks at ~1.6 and ~2 eV, which correspond to the optical gap energies of monolayer $MoSe_2$ and $WS_2$, respectively, indicating that perturbation to the electronic structure of individual monolayer is minimal as expected from the stacking disorder. The emission peaks are, however, slightly red–shifted and reduced in intensity in the heterostructure. For $MoSe_2$ emission, both trion and exciton components are observed, and the fraction of excitons is larger in the heterobilayer than in the monolayer $MoSe_2$ (Fig 1f, left), indicating de–doping of $MoSe_2$ layer upon equilibration with $WS_2$. For $WS_2$ emission, the predominant component is attributed to negative trion emission with negligible exciton emission, similar to observations in earlier reports[22–24] (Fig 1f, right). Thus, the heterostructure is an n–n junction with a slightly larger excess charge on the $WS_2$ layer due to charge redistribution upon contact. This behavior is consistent with the larger electron affinity of $WS_2$[25, 26].

The reduced PL intensity of the heterostructure can be attributed to an emergence of new decay channels. Given the type–II band alignment, the reduction in PL intensity is consistent with spontaneous photoinduced CT and formation of interlayer excitons[15, 27, 28]. The absence of low energy emission features due to radiative recombination of interlayer excitons may be attributed to the momentum–indirect interlayer band gap[29, 30]. The absence of the low energy interlayer emission peak is not unusual for twisted heterobilayers[14, 31]. The observation implies that the radiative recombination of interlayer excitons is a low yield process. This is likely considering the incommensurate stacking and indirect band gap of the system.

We show below that photoluminescence excitation (PLE) spectroscopy across exciton absorption resonances offers useful information on the coupling of the excited states in the heterobilayer. Figure 2a shows the color plot of PLE intensity measured at 78 K. In contrast to the monotonous PLE intensity map of individual monolayers[32], multiple peaks are evident in the excitation spectrum of both $MoSe_2$ and $WS_2$ A exciton emissions. For instance, the intensity of $MoSe_2$ A exciton emission is enhanced when the excitation is in resonance with the A, B, and C absorption features of $WS_2$ (Figure 2b). Similarly, the PLE intensity of $WS_2$ A exciton emission is slightly enhanced near the C peak of $MoSe_2$ (Figure 2c). The PLE results reveal 2s and 3s peaks of $WS_2$ A excitons at 2.18 and 2.25 eV, respectively (Figure 2c inset), coinciding with those reported for an isolated monolayer $WS_2$[33]. This indicates that the exciton binding energy for $WS_2$ is not altered due to hybridization with $MoSe_2$.

What is the origin of coupling between the excitation and emission states of the two layers? In a type–II heterostructure, it is natural to expect that photodoping influences the emission intensity of the layer accepting the charge. In order to examine the extent of this effect, we carefully monitored the intensity of trion and exciton emission of $MoSe_2$ at 15 K (Figure 2d and e). It can be seen that the exciton–trion ratio remains constant across the B exciton resonance of $WS_2$. That is, the PLE intensity modulation does not arise from changes in carrier densities. Thus, we rule out photodoing and conclude that ET is responsible for the emergence of resonance features in the PLE spectrum[34].

In order to estimate the ET and CT rates, we carefully study the enhancement factor ($\eta$), which we define here as the ratio of the integrated PL intensity for the heterostructure and isolated monolayer or $\eta = I^{het}/I^{mon}$. Large quenching ($\eta \ll 1$) implies rapid decay in exciton population due to newly introduced nonradiative relaxation channels. Note that when evaluating $\eta$, we take into account the effect arising from the differences in equilibrium doping in monolayer and heterobilayer using mass action model[35] (Supporting Note 4 and Figure S6).

Further, low energy emission due to localized states[36, 37] is excluded in this analysis. We show $\eta$ as a function of excitation energy for MoSe$_2$ A exciton emission normalized at the B exciton resonance of MoSe$_2$ in Figure 2f. The significance of the normalized $\eta$, denoted here as $\eta_M^0$ (where the subscript represents MoSe$_2$), is that $\eta_M^0 > 1$ indicates influx of exciton population (or energy) from WS$_2$. We find $\eta_M^0 > 1$ for all excitations above WS$_2$ A exciton energy, implying effective enhancement of PL for MoSe$_2$ by ET from WS$_2$. Figure 2g shows the enhancement factor for WS$_2$ A exciton emission normalized at the B exciton resonance of WS$_2$, denoted as $\eta_W^0$. Again, $\eta_W^0 > 1$ implies influx of exciton population from the MoSe$_2$ layer. We find that $\eta_W^0$ exceeds unity and peaks at MoSe$_2$ C resonance, which implies ET from MoSe$_2$ to WS$_2$ (Figure 2g). Since the band gap of MoSe$_2$ is significantly smaller than for WS$_2$, this observation suggests that the ET process competes with carrier thermalization process, which is also a fast process ($\leq$ 500 fs[38, 39]). Hot ET may be facilitated by the fact that excitation at MoSe$_2$ C peak resonance leads to spontaneous formation of temporarily stable indirect excitons[32]. The apparent contradiction that both layers appear to be enhanced at C resonance of MoSe$_2$ may be attributed to reduction in nonradiative decay rate in the heterobilayer (see Supporting Note 5 for details). In the following, we focus on the ET from WS$_2$ to MoSe$_2$.

Here, we use a model of single exciton dynamics to estimate the CT and ET rates from the enhancement factors and PL lifetimes (Figure S7) at different excitation energies. First, we consider excitation tuned to MoSe$_2$ B peak resonance where only MoSe$_2$ layer in the heterostructure is excited (Figure 3a). Assuming that the reduced PL intensity in the heterostructure can be solely attributed to interlayer electron transfer to WS$_2$, the enhancement factor for MoSe$_2$ A exciton emission $\eta_M$ (ref [1]) can be written as (see Supporting Note 6 for details):

$$\eta_M = \frac{I_M^{het}}{I_M^{mon}} = \frac{k_{r,M} + k_{nr,M}}{k_{r,M} + k_{nr,M} + k_{CT,M-W}}, \quad (1)$$

where $I_M^{mon}$ and $I_M^{het}$ are, respectively, the PL intensity of MoSe$_2$ for monolayer and heterostructure, $k_{r,M}$ and $k_{nr,M}$ are the radiative and nonradiative decay rates, and $k_{CT,M-W}$ is the rate of electron tunneling from MoSe$_2$ to WS$_2$. The subscript M represents MoSe$_2$. The values of $I_M^{mon}$ and $I_M^{het}$ are obtained from the steady−state PL measurements for excitation at the MoSe$_2$ B peak whereas $k_{r,M} + k_{nr,M}$ is obtained from PL lifetime measurements, Note that the resonance energy was determined from the differential reflection spectrum (Figure S5). Equation (1) assumes that radiative recombination rate and nonradiative decay rate of monolayer MoSe$_2$ remain unchanged when the supporting surface changes from SiO$_2$ to WS$_2$. However, since MoSe$_2$/WS$_2$ interface is expected to contain less trap states compared to MoSe$_2$/SiO$_2$ interface[33, 40, 41], it is likely that $k_{nr,M}$ is smaller for the heterobilayer. Thus, Equation (1) yields a lower bound for $k_{CT,M-W}$. It is worth noting that changes in dielectric and free carrier screening may also alter $k_r$. However, the unaltered exciton binding energy estimated from 2s and 3s peaks of WS$_2$ A excitons (Figure 2c inset) suggests that the spontaneous emission lifetime is only marginally affected by such effects. Based on these assumptions, we find the upper bound of CT time $(k_{CT,M-W})^{-1}$ to range between 0.6 and 3.6 ps at temperatures between 5 and 200 K, which are consistent with the previously reported subpicosecond values[14, 16].

We now estimate the rate of hole tunneling $k_{CT,W-M}$ and ET $k_{ET}$ (both from WS$_2$ to MoSe$_2$) independently from $k_{CT,M-W}$. For excitation at WS$_2$ B exciton resonance, WS$_2$ A exciton emission is quenched due to both ET and CT while MoSe$_2$ receives an influx of energy due to ET from WS$_2$ (Figure 3b). The enhancement factor of MoSe$_2$ emission $\eta_M$ can be expressed as:

$$\eta_M = \frac{(1-\beta_{CT,M\text{-}W})(A_W \beta_{ET} + A_M)}{A_M}, \qquad (2)$$

where $A_W$ and $A_M$ represent absorption by $WS_2$ and $MoSe_2$ layer, respectively, $\beta_{CT,M-W}$ is the efficiency of electron transfer from $MoSe_2$ to $WS_2$ and $\beta_{ET}$ is the ET efficiency for $WS_2$ (see Supporting Note 6 for details). On the other hand, the enhancement factor for $WS_2$ emission can be expressed as a function of $k_{CT,W-M}$:

$$\eta_W = \frac{I_W^{het}}{I_W^{mon}} = \frac{k_{r,W} + k_{nr,W}}{k_{r,W} + k_{nr,W} + k_{CT,W\text{-}M} + k_{ET}}. \qquad (3)$$

Here, we experimentally obtain $k_{r,W} + k_{nr,W}$ from PL lifetime measurements. Solving Equation (2) and (3) simultaneously, we obtain the two unknowns $k_{ET}$ and $k_{CT,W-M}$. Again, note that the above analysis yields a lower bound for $k_{ET}$ and $k_{CT,W-M}$. On the other hand, upper bound estimates can be obtained for $k_{nr,W}^{het} \to 0$ as discussed in Supporting Note 7 and 8. The upper bound values range between $(15 \text{ fs})^{-1}$ and $(45 \text{ fs})^{-1}$ for both CT and ET at temperatures between 5 and 200 K. Estimations based on changes in lifetime broadening[12] yielded values that fell between the upper and lower bounds, indicating consistency of our model.

The lower bound estimates of $k_{ET}$ and $k_{CT,W-M}$, and their corresponding efficiencies as a function of temperature are shown in Figure 4. We find the ET and CT rates to be of equal magnitude and weakly temperature dependent (Figure 4a). $k_{CT,M-W}$ and $\beta_{CT,M-W}$ were found to exhibit similar behaviors (Figure S8). At low temperatures, the ET time is within ~ 1 ps, which is about 3 orders of magnitude faster than the previously reported ET dynamics in coupled GaAs QWs separated by a thick (> 10 nm) barrier layer[3], and comparable to ET between single–walled carbon nanotubes[42,43] (typical Förster ET rates in various system are summarized in Table S1).

We now discuss the possible ET mechanisms at play in this system. These include dipole–dipole or near–field coupling (Förster–type), interlayer electron exchange (Dexter–type), and photon exchange (radiative). According to the theoretical analysis by Lyo[44], each mechanism manifests in a unique temperature–dependent ET rates. According to this analysis and experimental observations[44], Förster coupling is only weakly sensitive to temperature for small interlayer spacings[3,44]. As shown in Figure 4a, the temperature dependence of $k_{ET}$ agrees reasonably well with theoretical prediction for Förster ET between plane–wave 2D excitons[44]. The deviations at low temperatures may be attributed to localization of excitons in the $WS_2$ layer[44–46]. Photon–exchange ET is unlikely because the estimated radiative recombination lifetime of excitons in our system is significantly longer than the intrinsic lifetime of excitons in the light cone[46]. Thus majority of the excitons are outside the light cone due to thermal effects and can only interact nonradiatively. This is consistent with the classical theory that predicts strong near–field (nonradiative) coupling over far–field (radiative) coupling for short inter–exciton separations. Quantum electrodynamic treatment of TMD heterobilayers predicts electromagnetic ET rates that are also in subpicosecond time scales[46]. On the other hand, Dexter–type mechanism requires that both electrons and holes transfer from the donor layer to the acceptor layer. The type–II band alignment of our system dictates that electron transfer from $WS_2$ to $MoSe_2$ be thermally assisted. Since the temperature dependence of ET rates does not exhibit activated behavior (Figure 4a), we rule out this mechanism.

In molecular donor–acceptor systems, Förster ET requires spectral overlap in the excited states of the donor and acceptor species. In our heterobilayer system, A and B exciton resonances of MoSe$_2$ and WS$_2$ do not overlap with each other, suggesting that direct resonant excitation of 1s excitons via dipole–dipole interaction is unlikely. However, the WS$_2$ A exciton dipole can be in resonance with higher–order exciton states of A or B excitons in MoSe$_2$ (Figure 3). Coupling to free electron–hole pairs in MoSe$_2$ may also be possible but this is less likely considering the weak oscillator strength of free carriers[47, 48].

In order to further test the effect of ET, we conducted PLE measurements on a MoSe$_2$/hBN/WS$_2$ heterostack (Fig 5a and b) where the insulating hexagonal boron nitride (hBN) layer with a thickness of 6 nm is used is to suppress CT (see Figure S10 and Note 10). Figure 5c shows the PL spectra for MoSe$_2$ A exciton emission in monolayer MoSe$_2$, MoSe$_2$/WS$_2$ heterobilayer, and MoSe$_2$/hBN/WS$_2$ heterotrilayer excited in resonance with WS$_2$ A peak at 2.00 eV (PL spectra excited at 2.33 eV are shown in Figure S11 for reference). The emission is clearly enhanced in heterotrilayer in stark contrast to quenching observed in MoSe$_2$/WS$_2$ heterobilayer. This reveals that the effective excitation rate or absorption cross−section for MoSe$_2$ has increased in the presence of WS$_2$ due to ET. The PLE spectrum for MoSe$_2$ A exciton emission for MoSe$_2$/hBN/WS$_2$ heterotrilayer shown in Figure 5d shows an ET peak (W−A) for MoSe$_2$ emission, further evidencing optical pumping effect.

In summary, ET in van der Waals heterostructures of MoSe$_2$ and WS$_2$ monolayers were found to be highly efficient despite the type−II band alignment and competing ultrafast decay due to CT. Our results suggest that ET occurs via Förster coupling between ground excitons in WS$_2$ layer with upper–lying exciton states of MoSe$_2$ layer. Such strong dipole–dipole interactions are expected to be ubiquitous in heterostructures of other 2D semiconductors. The estimated ET rates in this system are some of the fastest among those reported for other nanostructure hybrid systems (Figure S12 and Table S1). We anticipate that the ability to sensitize 2D materials at specific resonances and manage energy transport at subnanometer length scales in intelligently designed heterostructures will enable novel approaches to enhancing photodetection, energy harvesting, and optical down conversion.

**Sample Preparation.**

The samples used in this study were monolayer MoSe$_2$ and WS$_2$ crystalline flakes. Single bulk crystals of MoSe$_2$ and WS$_2$ were grown by chemical vapor transport using iodine as the transport agent.[49] The crystals were mechanically exfoliated on the elastic polymer and transferred onto quartz substrates by stamping technique using silicone elastomer films to fabricate the heterostructure[50]. Finally, the heterostructure sample was annealed at 150 °C for 2 h and 200 °C for 1 h in vacuum (~ 10$^{-5}$ mbar pressure). The sample is characterized by PL, differential reflectance, Raman spectra, atomic force microscope, and PLE spectra. (see Figs. S1, S2, S3 and Note 1).

**Optical Measurements.**

The measurements of differential reflectance were performed using a tungsten–halogen lamp. The confocal micro–PL and PLE spectra under a back scattering geometry were obtained by monochromator and a supercontinuum light as an excitation source coupled to a monochromator. The excitation intensities for PL and PLE measurements were kept below 30 μW to avoid significant exciton–exciton annihilation[51, 52]. The measured spectral data were corrected for variations in the detection sensitivity with the correction factors obtained by using a standard tungsten–halogen lamp. Low temperature PL measurements were conducted for the samples in a liquid–helium–cooled cryostat.

ASSOCIATED CONTENT

Additional information including differential reflectance, Raman and PL spectra at room temperature, AFM, PLE spectra before and after annealing, calculated band structure, temperature dependence of differential reflectance and PL spectra, PL decay curves, decay rates and efficiencies, calculated energy transfer rates, Plots of energy transfer rate in various systems. (PDF)

AUTHOR INFORMATION
**Corresponding Author**
*Email: g.eda@nus.edu.sg

The authors declare no competing financial interest.

ACKNOWLEDGMENT

We would like to thank Prof. Farhan Rana for insightful discussions. We thank Lizhong Zhou for supporting optical measurements. G.E. acknowledges Singapore National Research Foundation for funding the research under NRF Research Fellowship (NRF−NRFF2011−02) and medium−sized centre programme. D.K., S.M. and K.M. are thankful for the financial support by a Grant−in−Aid for Scientific Research from the Japan Society for Promotion of Science (JSPS) (Grant No. 15J07423, 16K17485, 25400324, 26107522, 15K13500). Y.M. acknowledges support from Grant−in−Aid for Scientific Research 24681031, 15H05408, 15K13337 from JSPS, and by Precursory Research for Embryonic Science and Technology (PRESTO) Grant from Japan Science and Technology Agency (JST). A.C. and A.H.C.N. were supported by the NRF of Singapore (Grant number R−144−000−295−281). First−principles calculations were carried out on the CA2DM GRC high−performance computing facilities.

# Figures

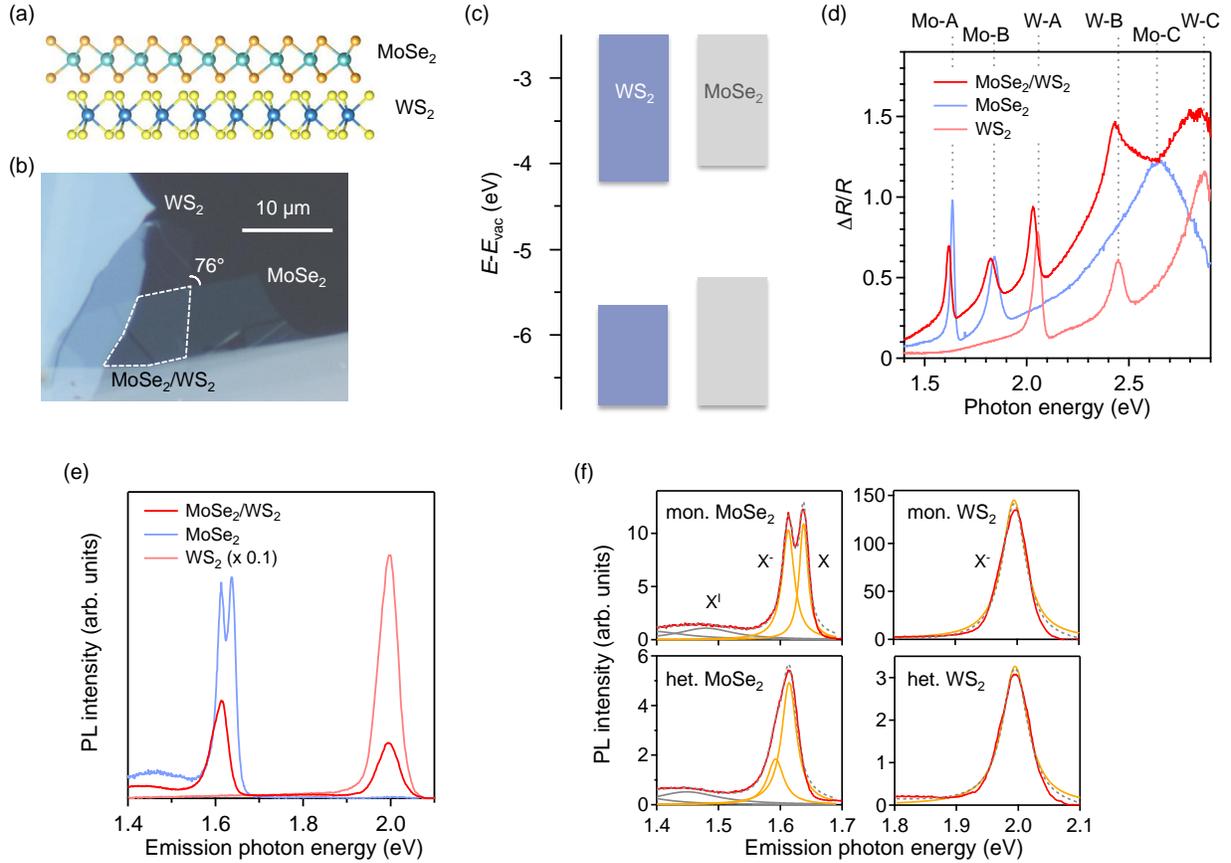

**Figure 1.** PL and differential reflectance spectra. (a) Schematic atomic structure of $MoSe_2/WS_2$ heterostructure. (b) Optical microscope image of the incommensurately stacked heterostructure (enclosed by a dashed line) on a quartz substrate. The scale bar corresponds to 10 μm. (c) Calculated band alignment of heterobilayer, exhibiting type−II alignment. (d) Differential reflectance spectra of heterobilayer and isolated monolayers at 78 K. The differential reflectance is defined as $\Delta R/R = (R_{S+Q} - R_Q)/R_Q$, where $R_{S+Q}$ and $R_Q$ are the reflected light intensities from the quartz substrate with and without the material, respectively[21, 49, 53]. (e) PL spectra of the heterostructure, isolated $MoSe_2$ and $WS_2$ monolayers measured at 78 K and at 2.44 eV excitation. (f) Deconvolution of PL spectra shown in (panel e) revealing exciton and trion components. The gray dashed lines are the sum of components, which are Lorentzian fits to exciton ($X^0$), trion ($X^-$) and impurity ($X^I$) emission peaks (orange solid lines).

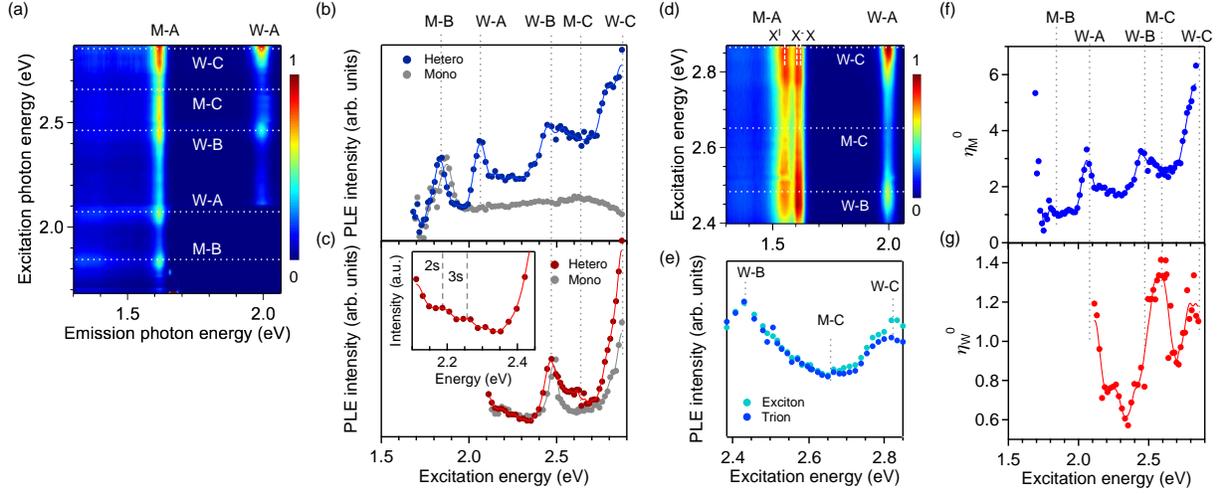

**Figure 2.** PLE spectra and enhancement factor. (a) PLE intensity map of $MoSe_2/WS_2$ heterostructure at 78 K where the color scale represent emission intensity. Exciton resonance energies of each layer are indicated by horizontal dotted lines. PLE spectra for monolayer and heterobilayer at (b) $MoSe_2$ A (M−A) and (c) $WS_2$ A (W−A) exciton emission energies at 1.61 and 1.98 eV, respectively. The inset shows PLE spectrum around 2s and 3s peaks of $WS_2$ A exciton in heterobilayer. The solid lines are guide to the eyes. (d) PLE intensity map of the heterostructure at 15 K. For $MoSe_2$ emission, the spectra clearly reveal three components: exciton ($X^0$); trion ($X^-$); and lower energy ($X^I$) emission. (e) PLE spectra for $MoSe_2$ emission. Normalized enhancement factor of (f) $MoSe_2$ ($\eta_M^0$) and (g) $WS_2$ ($\eta_W^0$) emission as a function of excitation energy. The spectra are normalized at the value corresponding to M−B and W−B excitations, respectively. The solid lines are guide to the eyes.

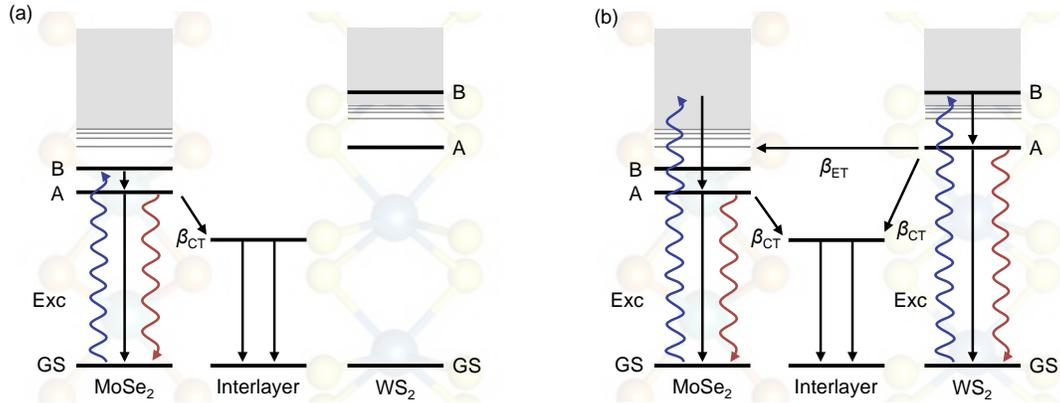

**Figure 3.** Exciton energy decay channels. Energy diagrams representing exciton relaxation channels in MoSe$_2$/WS$_2$ heterostructure for excitation at (a) MoSe$_2$ B resonance and (b) WS$_2$ B resonance. The wavy red and straight black arrows represent radiative and nonradiative decay processes, respectively. Thin horizontal lines above M−B and W−A levels represent higher order exciton states. $\beta_{CT}$ and $\beta_{ET}$ represent ICT and ET channels, respectively. The onset of the continuum states (gray box) is based on the previous reports[11, 47].

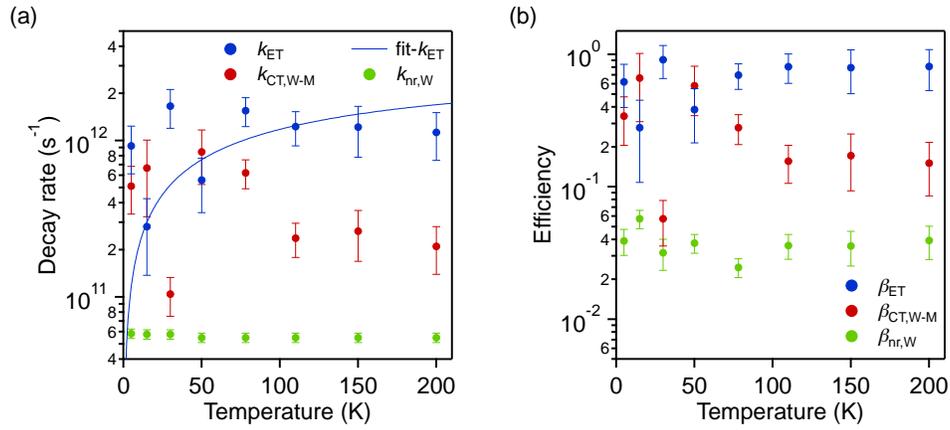

**Figure 4.** Decay rates and efficiencies. (a) ET, CT, and nonradiative decay rates of excitons in $WS_2$ as a function of temperature. The solid blue line is the fit based on the theoretical model by Lyo[44]. Details on the model are found in Supporting Note 9 and Fig, S9. CT represents hole transfer from $WS_2$ to $MoSe_2$ layer. (b) Efficiency for ET, CT and nonradiative decay as a function of temperature.

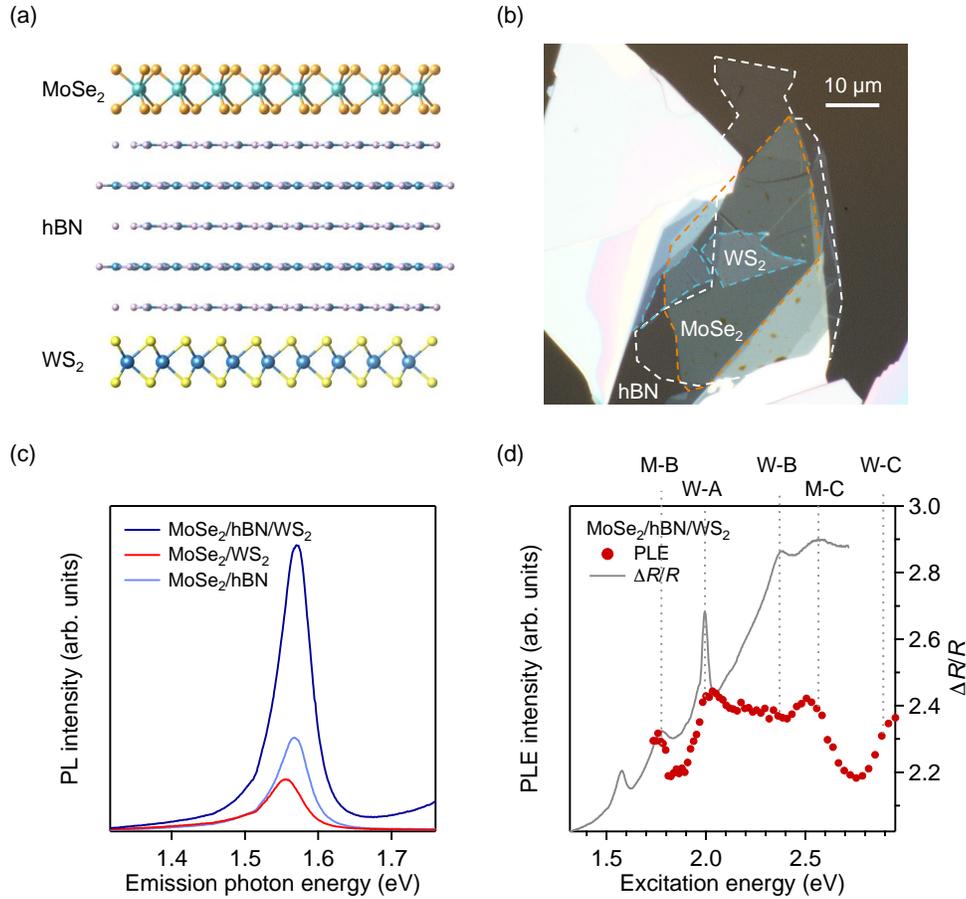

Figure 5. (a) Schematic atomic structure of $MoSe_2$/hBN/$WS_2$ heterotrilayer. (b) Micrograph of the heterotrilayer, where blue, white, and orange dashed lines respectively represent $WS_2$, hBN, and $MoSe_2$. (c) PL spectra for $MoSe_2$ emission in monolayer $MoSe_2$, $MoSe_2$/$WS_2$ heterobilayer, and $MoSe_2$/hBN/$WS_2$ heterotrilayer excited in resonance with $WS_2$ A at 2.00 eV at room temperature. (d) PLE spectra of band edge emissions for $MoSe_2$ in the heterotrilayer. The PLE spectra are normalized by the intensity of B peaks for band edge emission in $MoSe_2$. Differential reflectance spectra are also shown for comparison.

**Supporting Information**

# Evidence for Fast Interlayer Energy Transfer in MoSe$_2$/WS$_2$ Heterostructures

*Daichi Kozawa[a,b,c], Alexandra Carvalho[d,e], Ivan Verzhbitskiy[d,e], Francesco Giustiniano[d,e], Yuhei Miyauchi[a,f], Shinichiro Mouri[a], A. H. Castro Neto[d,e], Kazunari Matsuda[a], Goki Eda[\*,d,e,g]*

[a]Institute of Advanced Energy, Kyoto University, Gokasho, Uji, Kyoto, Japan 611-0011
[b]Department of Applied Physics, Waseda University, 3-4-1, Okubo, Shinjuku, Tokyo Japan 169-8555
[c]Department of Applied Physics, Nagoya University, Furo-cho, Chikusa-ku, Nagoya Japan 464-8603
[d]Department of Physics, National University of Singapore, 2 Science Drive 3, Singapore 117551
[e]Centre for Advanced 2D Materials, National University of Singapore, 6 Science Drive 2, Singapore 117546
[f]PRESTO, Japan Science and Technology Agency, 4-1-8 Honcho Kawaguchi, Saitama Japan 332-0012
[g]Department of Chemistry, National University of Singapore, 3 Science Drive 3, Singapore 117543

[\*]E-mail: g.eda@nus.edu.sg
[\*]Fax: +65 6777 6126

**Supporting Note 1.    Sample characterization**

The heterostructure was prepared by stamping technique[1,2] transferring the monolayers from elastic silicone polymer onto quartz substrates. We verified the number of layers for samples by differential reflectance, Raman and PL spectra (Figure. S1). The wavenumber of the Raman peaks agree with previously reported values for monolayer $MoSe_2$ and $WS_2$[3–6]. The heterostructure sample was annealed to enhance the interlayer coupling with decreasing the interlayer separation[7,8]. The AFM images of the heterostructure confirm that the two layers contact vertically (Figure S2). The annealing removes the bubble−like features on the hetero−region and improves the contact.

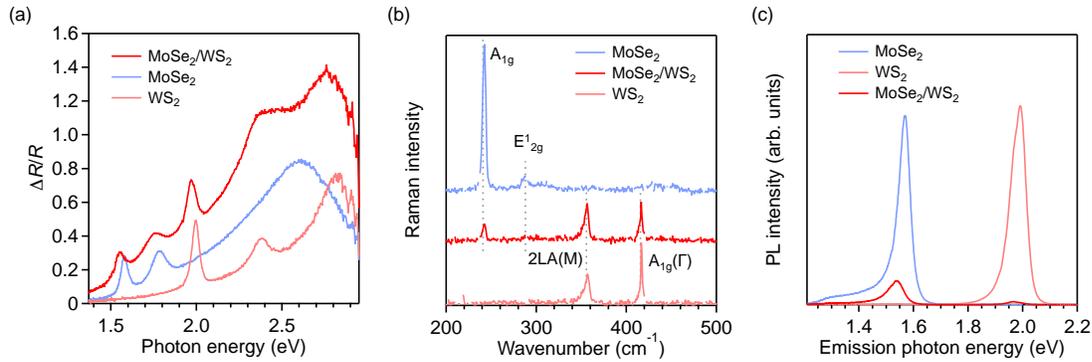

**Figure S1. Sample characterization by optical spectroscopy.** (a) Differential reflectance, (b) Raman and (c) PL spectra for $MoSe_2/WS_2$ heterostructure, isolated $MoSe_2$ and $WS_2$ monolayers at room temperature. In the Raman and PL measurements, 473 nm and 532 nm laser excitations were used, respectively. The Raman peaks are labelled for individual modes[9].

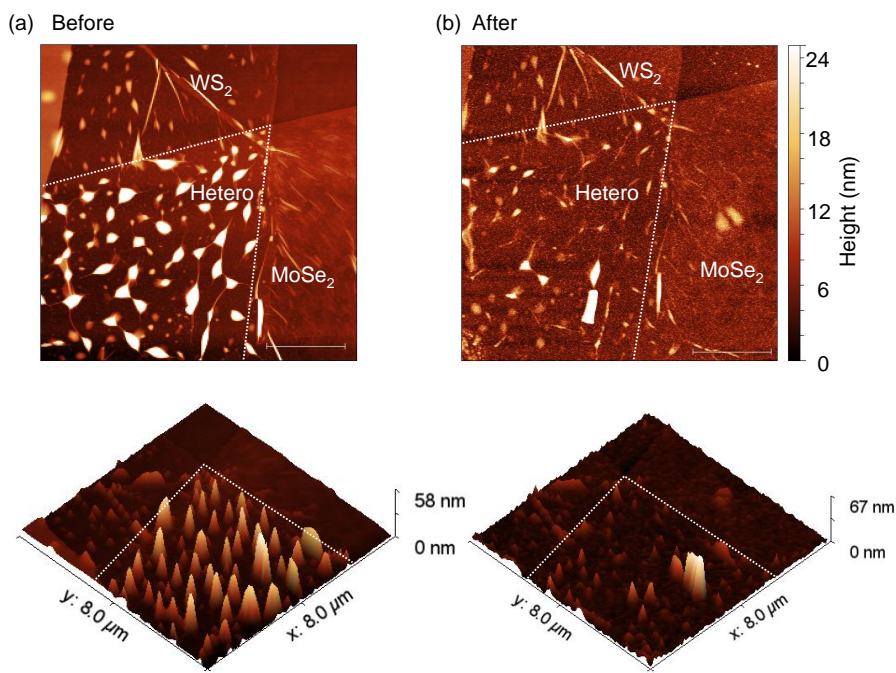

**Figure S2. Annealing effect on the surface morphology.** AFM and its three−dimensional images for (a) before and (b) after annealing $MoSe_2/WS_2$ heterostructure. The dotted lines indicate the boundary between hetero and isolated regions. After annealing, the polymer residues are significantly reduced to be the flat surface.

Figure S3 shows the PLE intensity map and spectra for heterobilayer before and after the annealing. Before the annealing, the PLE spectrum of this sample shows weaker energy transfer peak (W−A) compared to annealed sample which has significantly reduced bubbles. These data clearly show that the annealing enhance the interlayer coupling, and importantly, the good contact is required to emerge the energy transfer (ET) and it does not preferentially occur at the bubbles.

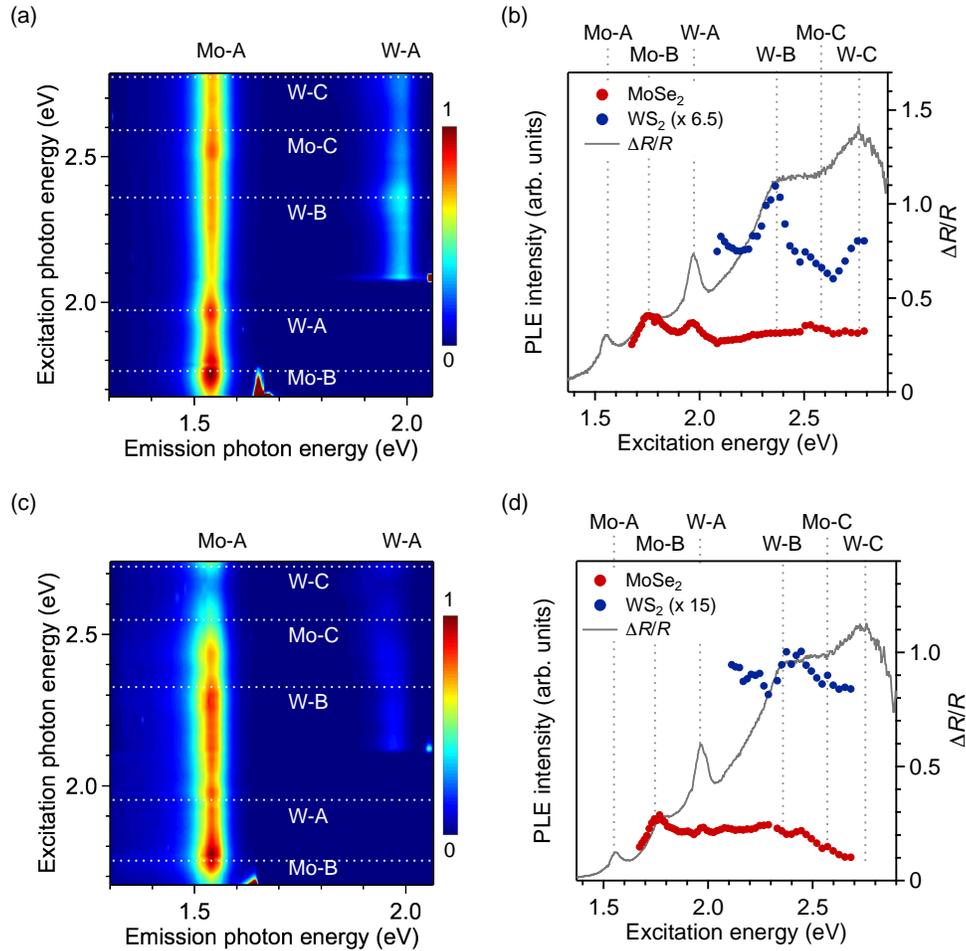

**Figure S3. PLE spectra for MoSe$_2$ and WS$_2$ emission in the heterostructure at room temperature.** (a) PLE intensity map of MoSe$_2$/WS$_2$ heterostructure measured at room temperature. Typical resonant energies are indicated by dotted lines. (b) PLE spectra of band edge emissions for MoSe$_2$ and WS$_2$ in heterostructure. The PLE spectra are normalized by the intensity of B peaks for band edge emission in MoSe$_2$ and WS$_2$, respectively. Differential reflectance spectra are also shown for comparison. (c) PLE intensity map and (d) spectra for the heterostructure before the annealing. The PLE spectra for WS$_2$ emissions are vertically displaced with a certain offset.

Both charge and energy transfers are most efficient where the two layers are in physical contact. Interlayer tunneling probability exponentially decays with layer separation. Dipole−dipole energy transfer rate is expected to decay by $1/d^4$. Further more, imperfections in our heterostructures such as wrinkles and bubbles are present only in small areal fractions (<10%). Thus, our measured signals are representative of the effects that are occurring at the interface of clean heterobilayer region, rather than at the structural imperfections.

**Supporting Note 2.    Calculation of band structure for the type−II heterobilayer**

We conducted a series of DFT calculations for the semiconductor TMDs family using the open source code QUANTUM ESPRESSO[10]. We used norm−conserving, fully relativistic pseudopotentials with nonlinear core−correction and spin−orbit information to describe the ion cores. The pseudopotentials used were either obtained from the QUANTUM ESPRESSO distribution or produced using the ATOMIC code by A. Dal Corso. The exchange correlation energy was described by the generalized gradient approximation (GGA), in the scheme proposed by Perdew, Burke, and Ernzerhof (PBE)[11]. The integrations over the Brillouin zone were performed using a scheme proposed by Monkhorst−Pack[12] for all calculations. The energy cutoff was 30 Ry. The iso−surface of charge density is projected on a real−space in a similar way to the previous report of Ref. [13].

To examine the relaxation of electron−hole pairs, we compute the electronic structure of the heterostructure with finite stacking angle between the two layers as shown in Figure S4a. The stacking of $MoSe_2/WS_2$ lattices orients with an angle of 16.1 degrees in this calculation[14] and exhibits consequent displacement of K/K' points for $MoSe_2$ and $WS_2$ in the Brillouin zone (Figure S4b). Figure S4c and d shows the band structure of $MoSe_2/WS_2$ heterostructure, in which main features agree with a previous report[14–16]. In TMDs bilayer, the mixing around Γ points pushes up the valence band, which leads to make the gap indirect[14, 17, 18] (Figure S4c and d) with type−II band alignment (Figure S4e).

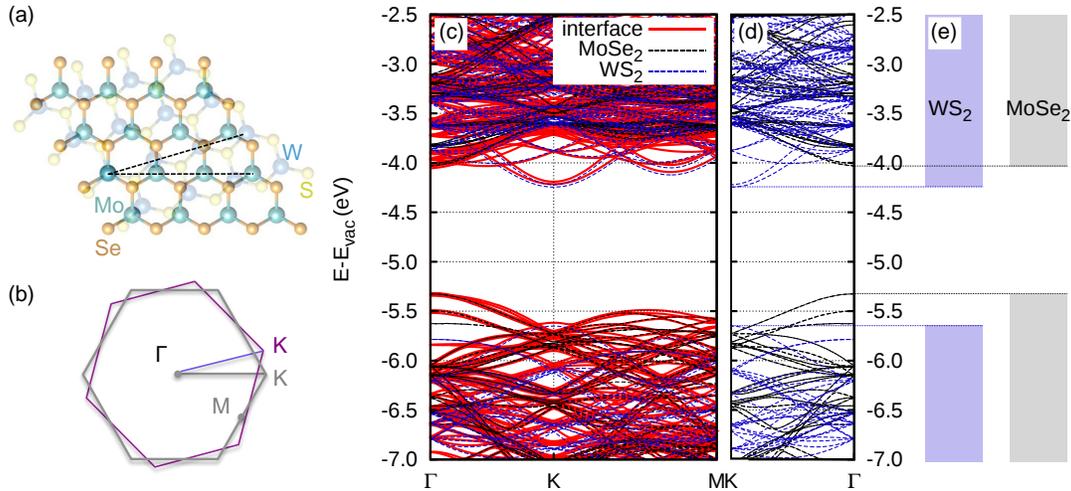

**Figure S4. Calculation of electronic structure.** (a) Top view of $MoSe_2/WS_2$ heterobilayer. (b) Brillouin zones for stacking $MoSe_2$ and $WS_2$ monolayers with finite angle−stacking, where highly symmetric points are labelled in the Brillouin zones for individual $MoSe_2$ and $WS_2$ monolayers. (c) Energy band structure of $MoSe_2/WS_2$ heterostructure (solid lines) and its components from $WS_2$ and $MoSe_2$ (dashed lines). (d) Energy band structure nearby band edge of $WS_2$ and $MoSe_2$. (e) Energy band alignment highlighting band edge of $WS_2$ and $MoSe_2$. The dotted horizontal lines indicate the energy of the band edges.

The band line−up shown in Figure 1c is obtained for an ideal bilayer heterostructure, consisting of two intrinsic, perfectly crystalline moieties. In the real system, both $MoSe_2$ and $WS_2$ contain defects unintentionally introduced, which confer doping to the heterostructure.

However, in the case of bilayer heterosystems with moderate doping level, it is reasonable to assume that the doping does not alter significantly the band alignment. Note that this is different from 3D systems, where the doped 'bulk' far from the junction provides an ideal reservoir of carriers with fixed Fermi level. In monolayer systems, the number of doping carriers per unit area is finite and its equilibration across the interface does not significantly change the interface voltage. The interpretation of bilayer and few layer heterojunction experiments so far therefore suggests that the best approximation is to obtain the band alignment of the interface from aligning the bands of the two moieties with respect to the vacuum level. This is the approach that we adopted.

As for the difference between the experimental optical bandgaps and the calculated bandgaps, it originates from different sources. One is that the experimental feature energies correspond to the optical gaps, which are excitonic gaps. The bands represented in Figure 1c are one−electron energy levels (which correspond approximately to the transport bandgap). Further, the density functional theory (DFT) calculations are known to underestimate the transport bandgap. Moreover, in the heterostructure there is some small misfit strain, which is an additional, but minor, source of error.

**Supporting Note 3.    Effect of doping on band alignment**

Isolated monolayers and heterobilayers show different levels of doping. We estimate this change in the doping density to be $< 10^{12}$ cm$^{-2}$ from the changes in the trion/exciton intensity ratio. Such change in the doping density is not expected to lead to a significant band renormalization effect as discussed in a recent study[19]. Charge transfer upon stacking leads to respective shift in the Fermi level but this should not affect the band alignment.

Formation of 2D p−n junction also does not affect the band alignment. Lee et al. reports that holes in p−doped WSe$_2$ and electrons in n−doped MoS$_2$ are not depleted in MoS$_2$/WSe$_2$ junction at a forward bias[20]. Even at zero bias, the junction only has a slight depletion region. In our system, both MoSe$_2$ and WS$_2$ are n−type, which likely leads to the lack of depletion. In the microbeam X−ray photoelectron spectroscopy and scanning tunneling microscopy/spectroscopy[21] also supports the type−II band alignment in MoS$_2$/WSe$_2$ heterobilayer. Their density functional theory agrees with their experimental results within 0.11 eV departures. We thus confirm that the band alignment of our MoSe$_2$/WS$_2$ is type−II configuration, as our DFT calculation suggests.

**Supporting Note 4.    Mass action model**

The difference in the charge density of monolayer and heterostructure affects the PL intensity. We used the mass action model[22] to correct the PL intensity for MoSe$_2$ emission to cancel the enhancement effect by the charge dedoping in the heterobilayer. Here X$^0$, X$^-$ and e$^-$ denote exciton, trion and free electrons. The formation of trion, $X^0 + e^- \rightarrow X^-$, can be described by a standard mass action model[22, 23]:

$$\frac{n_X n_e}{n_{X^-}} = \left(\frac{4 m_X m_e}{\pi \hbar^2 m_{X^-}}\right) k_B T \exp\left(-\frac{E_T}{k_B T}\right) \qquad (1)$$

where $n_X$, $n_e$ and $n_{X^-}$ are population of exciton, free electrons and trion, $m_X$, $m_e$, $m_{X^-}$ are the effective mass of exciton, electron and trion, $T$ is the temperature, $k_B$ is Boltzmann constant, $E_T$ is the trion binding energy. The above equation has been shown to fit the experimentally observed trend[22] Thus, from our experimentally obtained ratio of trion and

S6

neutral exciton emission peaks, the neutral exciton emission intensity of a dedoped sample can be estimated. We used this intensity to calculate the enhancement factor.

**Supporting Note 5. Hot exciton energy transfer**

The hot ET may be facilitated by the fact that excitation at the C peak resonance leads to spontaneous formation of temporarily stable indirect excitons[24]. The apparent contradiction that both layers appear to be enhanced at C resonance of MoSe$_2$ may be attributed to substrate effect. As we reported previously[24], the emission intensity at C peak excitation is sensitive to nonradiative decay rate. For instance, when a monolayer WS$_2$ is measured on hBN instead of on a SiO$_2$ surface, the overall QY is enhanced and the excitation energy dependence of QY is also altered. A similar effect comes into play in a heterobilayer. Since the MoSe$_2$ layer sits on the WS$_2$ layer in the heterostructure, it is possible that nonradiative decay rates are decreased compared to MoSe$_2$ directly sitting on quartz substrate.

**Supporting Note 6. Estimation of charge and energy transfer rate**

To evaluate the magnitude of the PL quenching, we introduce rate equation of the exciton dynamics. The photo−excitation creates exciton in MoSe$_2$ and WS$_2$ with generation factor $G_M$ and $G_W$, respectively. Excitons in the individual monolayer decay in radiative and nonradiative process with the rate of $k_r$ and $k_{nr}$ as an internal process. We treat the rate equation of the A−exciton population in monolayer $N_M^{mon}$ and $N_W^{mon}$:

$$\frac{dN_M^{mon}}{dt} = G_M - (k_{r,M} + k_{nr,M})N_M^{mon} \quad \text{and} \quad \frac{dN_W^{mon}}{dt} = G_W - (k_{r,W} + k_{nr,W})N_W^{mon}. \quad (2)$$

The population of exciton can be derived from the steady−state solutions and expressed as

$$N_M^{mon} = \frac{G_M}{k_{r,M} + k_{nr,M}} \quad \text{and} \quad N_W^{mon} = \frac{G_W}{k_{r,W} + k_{nr,W}}. \quad (3)$$

The emission intensity is proportional to the radiative decay rate as follows:

$$I_M^{mon} = \frac{G_M k_{r,M}}{k_{r,M} + k_{nr,M}} \quad \text{and} \quad I_W^{mon} = \frac{G_W k_{r,W}}{k_{r,W} + k_{nr,W}}, \quad (4)$$

where $I_M^{mon}$ and $I_W^{mon}$ are the emission intensity for monolayer MoSe$_2$ and WS$_2$. Note that the resonance energy was determined from the differential reflection spectrum (Figure S5).

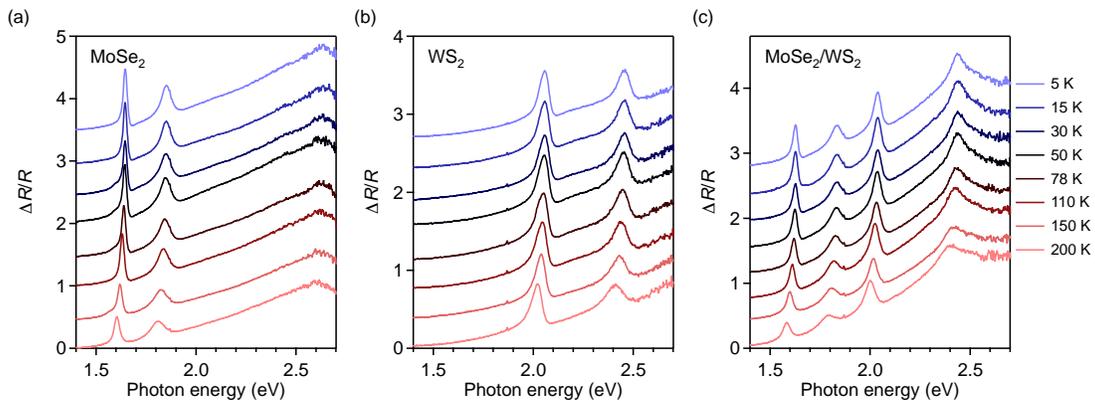

**Figure S5. Temperature dependence of differential reflectance spectra with various sample temperatures for** (a) MoSe$_2$, (b) WS$_2$ monolayers and (c) MoSe$_2$/WS$_2$ heterostructure.

With MoSe$_2$−B resonant excitation to heterostructure, only MoSe$_2$ is excited. The decay channel to the interlayer exciton through charge transfer is incorporated into the decay in monolayer MoSe$_2$ (Figure 3a). Then, the rate equation for MoSe$_2$ emission in heterostructure is expressed as

$$\frac{dN_M^{het}}{dt} = G_M - \left(k_{r,M} + k_{nr,M} + k_{CT,M\text{-}W}\right) N_M^{het} \quad (5)$$

where the superscript, *het*, represents heterostructure and $k_{CT,M–W}$ is electron transfer rate from MoSe$_2$ to WS$_2$. The PL intensity of heterostructure MoSe$_2$ emission is expressed as

$$I_M^{het} = \frac{G_M k_{r,M}}{k_{r,M} + k_{nr,M} + k_{CT,M\text{-}W}} \quad (6)$$

Then we can obtain the charge transfer rate from the ratio of the PL intensity as enhancement factor of MoSe$_2$ emission $\eta_M$:[25]

$$\eta_M = \frac{I_M^{het}}{I_M^{mon}} = \frac{k_{r,M} + k_{nr,M}}{k_{r,M} + k_{nr,M} + k_{CT,M\text{-}W}} \quad (7)$$

where $k_{CT}$ is unknown and all else parameters are experimentally obtainable. The values of the PL intensity are obtained in the static PL measurements (Figure S6), and those of $k_{r,M} + k_{nr,M}$ correspond to the PL lifetime in the time−resolved PL measurements (Figure S7).

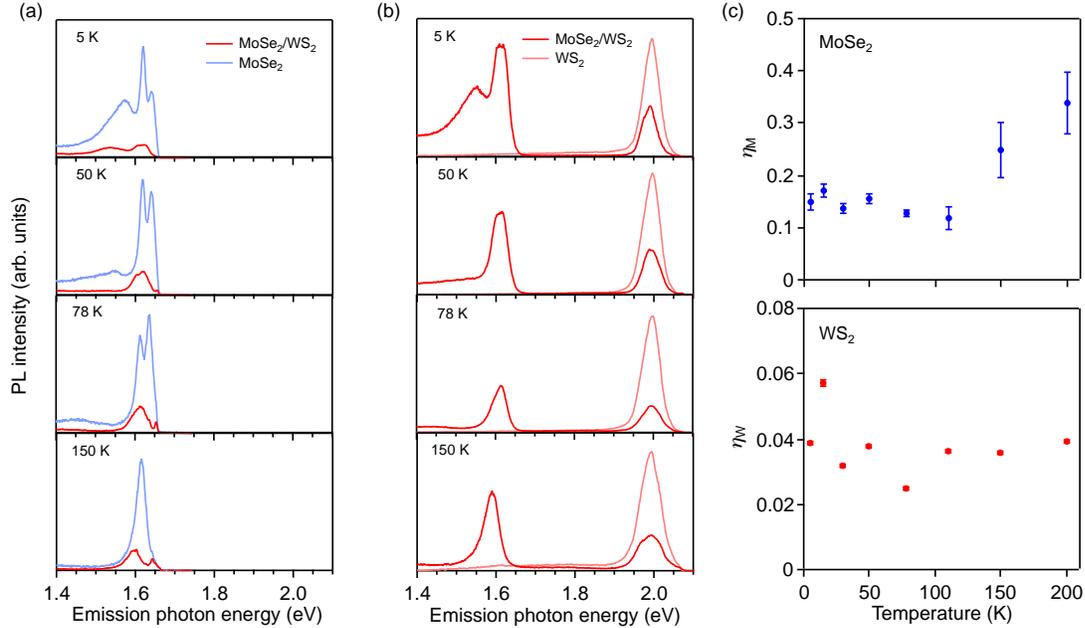

**Figure S6. Temperature dependence on PL quenching.** PL spectra for (a) MoSe$_2$ and (b) WS$_2$ emission in the monolayer and heterostructure at various temperatures. For MoSe$_2$ emission, B peak of MoSe$_2$ is excited, and for WS$_2$ emission, B peak of WS$_2$ is excited. (c) The enhancement factor through formation of the heterostructure for MoSe$_2$ (top) and WS$_2$ (bottom) emission as a function of temperature.

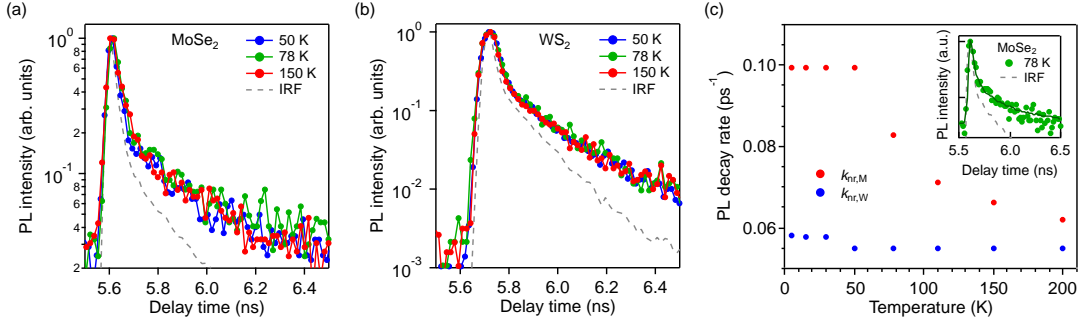

**Figure S7. Temperature dependence on PL decay.** PL decay curves for monolayer (a) MoSe$_2$ and (b) WS$_2$ emission at various temperatures with excitation of MoSe$_2$−B and WS$_2$−B peaks, respectively. The broken line shows the IRF corresponding to the excitation. (c) PL decay rate for MoSe$_2$ and WS$_2$ emission as a function of temperature. The inset displays the example of fitting curve with a convolution of IRF and double exponential model decay for MoSe$_2$ emission at 78 K, where the fast component is almost unity fraction.

For WS$_2$ B excitation resonance, the population of exciton in MoSe$_2$ is described as

$$N_M^{mon} \propto A_M \quad (8)$$

and

$$N_M^{het} \propto A_M + A_W \beta_{ET}(1-\beta_{CT,M-W}) - A_M \beta_{CT,M-W} \quad (9)$$

where $\beta_{CT,M-W}$ is the efficiency of electron transfer from MoSe$_2$ to WS$_2$ $\beta_{CT,M-W} = \dfrac{k_{CT,M-W}}{k_{r,M} + k_{nr,M} + k_{CT,M-W}}$, the first term is equivalent of absorbance in individual monolayer MoSe$_2$, the second is the inflow from WS$_2$ by the ET, and the third is the outflow by the charge transfer to WS$_2$ layer. Then the enhancement factor is

$$\eta_M = \frac{N_M^{het}}{N_M^{mon}} = \frac{(A_M + A_W \beta_{ET})(1-\beta_{CT,M-W})}{A_M} \quad (10)$$

where $\beta_{ET}$ is the efficiency of ET from WS$_2$ to MoSe$_2$ $\beta_{ET} = \dfrac{k_{ET}}{k_{nr,W} + k_{CT,W-M} + k_{ET}}$. We are able to obtain the ET rate $k_{ET}$ from the enhancement factor of WS$_2$ emission $\eta_W$, showing the following manner. The rate equation of WS$_2$ emission in heterostructure with an excitation at WS$_2$−B

$$\frac{dN_W^{het}}{dt} = G_W - (k_{r,W} + k_{nr,W} + k_{CT,W-M} + k_{ET})N_W^{het} \quad (11)$$

where the decay channel of exciton consists of the radiative, nonradiative decay, CT and ET (Figure 3b). Then the enhancement factor or the ratio of WS$_2$ emission in monolayer and heterostructure can be expressed as

$$\eta_W = \frac{I_W^{het}}{I_W^{mon}} = \frac{k_{r,W} + k_{nr,W}}{k_{r,W} + k_{nr,W} + k_{CT,W-M} + k_{ET}} \quad (12)$$

Solving the simultaneous equation of Equations (10) and (12), we obtain $k_{ET}$ and $k_{CT,W-M}$. The lower bound estimates of $k_{ET}$ and $k_{CT,W-M}$, $k_{CT,M-W}$ and their corresponding efficiencies as a function of temperature are shown in Figure 4 and Figure S8.

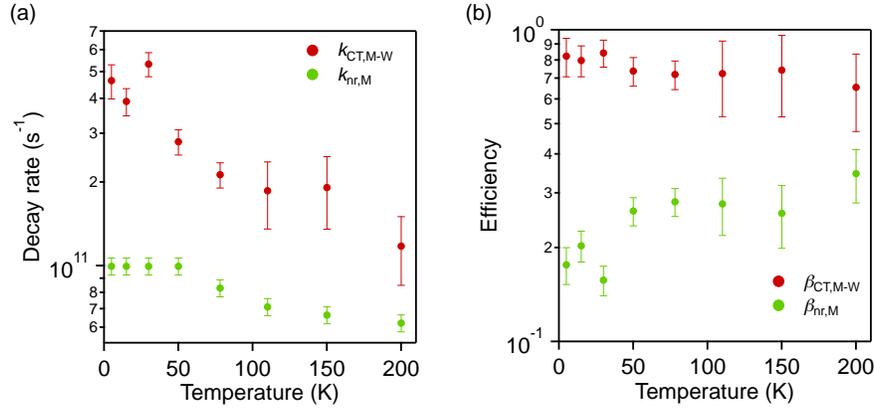

**Figure S8. Decay rates and efficiencies.** (a) Energy and charge transfer rate of exciton in $MoSe_2$ as a function of temperature. The subscripts M and W indicate $MoSe_2$ and $WS_2$. CT represents electron transfer from $MoSe_2$ to $WS_2$ layer (b) Efficiency for CT and nonradiative decay as a function of temperature.

**Supporting Note 7.    Estimated error on interlayer charge and energy transfer rate**

Now we examine possible changes in the nonradiative decay rate. Earlier studies have shown that monolayer $MoS_2$ and other similar materials exhibit up to 15 times increase in photoluminescence intensity when the substrate is changed from $SiO_2/Si$ to hexagonal boron nitride (hBN). This is most likely due to low density of nonradiative recombination centers on the surface of hBN, which is virtually free of dangling bonds and charge impurities. In such cases, one can conclude that the nonradiative decay rate has decreased by nearly an order of magnitude. The surface of $WS_2$ on which $MoSe_2$ is placed is of similar nature as that of hBN in that the density of structural defects is low. Thus, it is reasonable to assume that the nonradiative recombination rate of $MoSe_2$ on $WS_2$ is lower than that on quartz. Based on the early experimental reports, the magnitude of this change is expected to be less than a factor of 15.

In the main text, we assumed the same $k_r$ and $k_{nr}$ for the isolated monolayer and heterobilayer. However, while $k_r$ is not strongly affected based on earlier work on the dielectric effect[26, 27], $k_{nr}$ can decrease when a sample is placed on a cleaner subs[28]trate or interfacial trap states are removed[29–31]. Considering change in $k_{nr,M}$ to $k_{nr,M}^{het} = \xi_M k_{nr,M}$, where correction factor for $k_{nr,M}$ is $\xi_M < 1$. i.e. we assume that van der Waals interface is "better" than air/TMD or $SiO_2$/TMD because of no dangling bonds and reduced surface contamination. Then Equation (6) can be described as

$$\eta_M = \frac{I_M^{het}}{I_M^{mon}} = \frac{k_{r,M} + k_{nr,M}}{k_{r,M}^{het} + \xi_M k_{nr,M}^{het} + k_{CT,M-W}^{het}} . \quad (13)$$

where the CT rate $k_{CT,M-W}$ is modified to be $k_{CT,M-W}^{het}$, and we obtain

$$k_{\text{CT,M-W}}^{\text{het}} = \frac{k_{r,M}(1-\eta_M) + k_{nr,M}(1-\xi_M\eta_M)}{\eta_M}. \tag{14}$$

This equation tell us that $k_{\text{CT,M-W}}^{\text{het}} > k_{\text{CT,M-W}}$ since $\xi_M < 1$. We evaluate the correction factor to $k_{\text{CT,M}}$ in ratio given by

$$\frac{k_{\text{CT,M-W}}^{\text{het}}}{k_{\text{CT,M-W}}} = \frac{k_{r,M}(1-\eta_M) + k_{nr,M}(1-\xi_M\eta_M)}{k_{r,M}(1-\eta_M) + k_{nr,M}(1-\eta_M)}. \tag{15}$$

Assuming $k_r \ll k_{nr}$, the correction factor for $k_{\text{CT,M-W}}$ can be written as

$$\frac{k_{\text{CT,M-W}}^{\text{het}}}{k_{\text{CT,M-W}}} = \frac{1-\xi_M\eta_M}{1-\eta_M}. \tag{16}$$

Even in the extreme case of $\xi_M \to 0$ under the distinct improvement of interface, this ratio yields 1.1−1.5 of the correction when $\eta_M = 0.12$−0.34 (Figure S6c), respectively. In our system, the quenching is dominant over any enhancement factors. Thus, the error due to our assumption $\xi_M = 1$ is less than a factor of 1.5.

Similarly, considering change in $k_{nr,W}$ to $k_{nr,W}^{\text{het}} = \xi_W k_{nr,W}$, where correction factor for $k_{nr,W}$ is $\xi_W < 1$, we substitute this for Equations (10) and (12):

$$\eta_M = \frac{I_M^{\text{het}}}{I_M^{\text{mon}}} = \frac{(A_M + A_W\beta_{\text{ET}}^{\text{het}})(1-\beta_{\text{CT,M-W}}^{\text{het}})}{A_M}, \tag{17}$$

and

$$\eta_W = \frac{I_W^{\text{het}}}{I_W^{\text{mon}}} = \frac{k_{r,W} + k_{nr,W}}{k_{r,W}^{\text{het}} + \xi_W k_{nr,W}^{\text{het}} + k_{\text{CT,W-M}}^{\text{het}} + k_{\text{ET}}^{\text{het}}}. \tag{18}$$

Then we evaluate the correction factor to $k_{\text{CT,W-M}}$ and $k_{\text{ET}}$ in ratio

$$\frac{k_{\text{CT,W-M}}^{\text{het}}}{k_{\text{CT,W-M}}} = \frac{(\beta_{\text{CT,M-W}} - 1)\left[A_W\xi_W(\beta_{\text{CT,M-W}}^{\text{het}} - 1) - A_M\eta_M - (A_M + A_W)(\beta_{\text{CT,M-W}}^{\text{het}} - 1)\right]}{(\beta_{\text{CT,M-W}}^{\text{het}} - 1)\left[A_W(\beta_{\text{CT,M-W}} - 1) - A_M\eta_M - (A_M + A_W)(\beta_{\text{CT,M-W}} - 1)\right]} \tag{19}$$

and

$$\frac{k_{\text{ET}}^{\text{het}}}{k_{\text{ET}}} = \frac{(\beta_{\text{CT,M-W}} - 1)(\eta_M + \beta_{\text{CT,M-W}}^{\text{het}} - 1)}{(\beta_{\text{CT,M-W}}^{\text{het}} - 1)(\eta_M + \beta_{\text{CT,M-W}} - 1)}, \tag{20}$$

respectively. The upper bound values range between $(15 \text{ fs})^{-1}$ and $(45 \text{ fs})^{-1}$ for both CT and ET at temperatures between 5 and 200 K. Taken into account that the CT takes place within 50 fs in the previously reported time−domain measurements[32], the estimated error for CT rate is compatible with them.

**Supporting Note 8.  Temperature dependence of nonradiative decay rate**

The temperature dependence of the nonradiative decay rates depends on the recombination mechanism and there is clearly a scatter in the reported values[33-38]. The components of nonradiative decay in the heterobilayer are broken down into intrinsic, surface impurity−induced, and surface trap−induced nonradiative decays. For intrinsic nonradiative decay, the decay rate decreases with temperature[34], which we observed in the decay in isolated monolayer $MoSe_2$. For surface impurity−induced nonradiative decay, the impurities changes the doping level[39] and alter the initial trion formation. The fraction of trion decreases with temperature, which lead to decrease decay rate with temperature, since the decay rate for trion is ~1.8 times smaller than exciton (exciton for 14 ps, trion for 25 ps in monolayer $WSe_2$)[35].

For Previous work on $MoS_2$ nano−clusters presented the that surface traps consist of shallow and deep traps[40], where excitons are dominantly trapped on by the surface trapped states with fast decay of 2−4 ps[33]. These trapping of exciton are observed even at room temperature in the pump−probe measurements[33]. Thus we can perhaps attribute the observation that the nonradiave rates in the isolated monolayer $WS_2$ are independent on temperature to the fast trapping of exciton.

**Supporting Note 9.    Theoretical calculation of Förster energy transfer rate**

We simulate the temperature dependence of Förster energy transfer between separated quantum wells based on dipole−coupling model[41]. A plane−wave exciton in donor decays into a free electron−hole pairs in acceptor with rate $k_{ET}$. Then we have

$$k_{ET} = k_{ET}^0(R_T) g\left(\frac{d}{R_T}\right), \quad (21)$$

where $d$ (= 0.8 nm) is the separation of the layers and

$$g(t) = \int_0^\infty x^3 \exp(-x^2)\exp(-2tx)S(tx)^2 \, dx, \quad (22)$$

where $x$ is an integration variable. In Equation (22), $S$ is a function to estimate the magnitude of average electron and hole distribution and equals

$$S(t) = \frac{\sinh(tb_1/2d)}{tb_1/2d}\frac{\sinh(tb_2/2d)}{tb_2/2d}, \quad (23)$$

where $b_1$ and $b_2$ ($b_1 = b_2 = 0.8$ nm) are thickness of donor and acceptor layer.

$$k_{ET}^0(R_T) = \frac{32\pi\mu}{\hbar^3}\left(\frac{e^2 D_1 D_2}{\kappa a_B R_T}\right)^2 = \frac{c}{R_T^2}, \quad (24)$$

where $\mu$ is reduced exciton mass, $\hbar$ is reduced Planck constant, $D_1$ and $D_2$ are the dipole moment of donor and acceptor layer, $\kappa$ is average dielectric constant, $a_B$ is Bohr radius of exciton. However we assign the temperature−independent term in this equation as a fitting coefficient $c$. $R_T$ is the localization radius of exciton:

$$R_T = \sqrt{\frac{\hbar^2}{2m_x k_B T}}, \quad (25)$$

where $m_x$ (= $0.89m_0$, $m_0$ is electron mass[42]) is exciton mass. Substituting Equations (22)−(25) into (21), we obtain the ET rate. By plotting the $k_{ET}$ as a functions of $T$, we can depict the fitting curve shown in Figure 4a. We also obtain the contour plots of $k_{ET}$ with various $d$ and $T$ as shown in Figure S9.

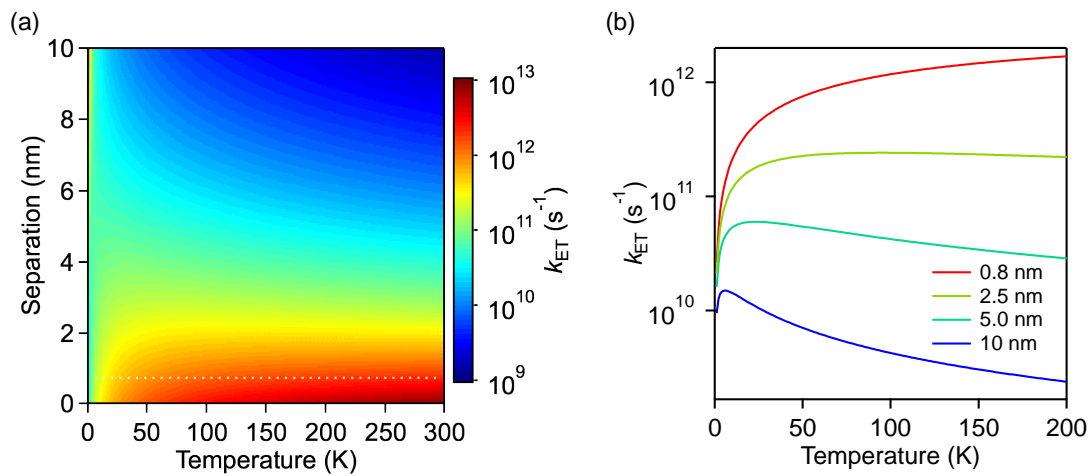

**Figure S9. Plots of Förster energy transfer model.** (a) 2D map of calculated ET rate as functions of separation and temperature. The dotted line corresponds to the applied separation of heterobilayer in the main text. (b) The profile of the map at 0.8, 2.5, 5.0, and 10 nm separation.

**Supporting Note 10.    Insertion of hBN spacer layer**

In principle, CT cannot be completely suppressed because there is always a finite probability of tunneling for charges. However the absence of quenching clearly indicates that CT is far less efficient compared to ET. Absence of major charge transfer is also evident from the $MoSe_2$ emission of heterotrilayer and $MoSe_2$/hBN bilayer (Figure S10). The emission spectra are effectively identical, indicating that no charge transfer has occurred upon equilibration in contrast to the case of $MoSe_2$ placed in direct contact with $WS_2$. This result suggests that interlayer charge transfer is largely suppressed due to the insulating hBN layer.

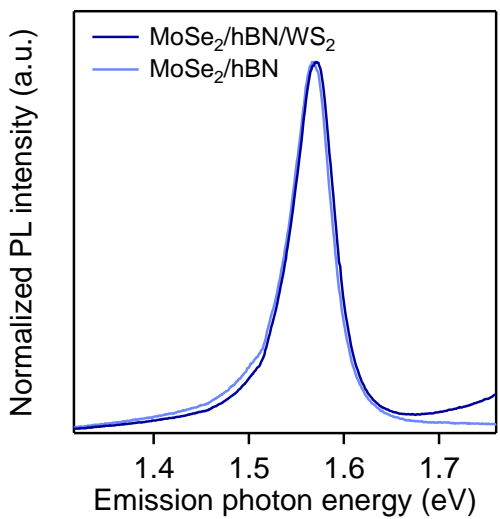

**Figure S10. PL spectra for heterotrilayer and $MoSe_2$/hBN** excited at 2.33 eV at room temperature.

Figure S11 is PL spectra for the MoSe$_2$/hBN/WS$_2$ heterotrilayer, MoSe$_2$/WS$_2$ heterobilayer, MoSe$_2$, WS$_2$ monolayers excited at 2.33 eV corresponding to WS$_2$−B resonance. In the heterotrilayer, MoSe$_2$ emission is enhanced compared to isolated monolayer and heterobilayer, while WS$_2$ emission is quenched. These results show that the effective excitation rate or absorption cross−section for MoSe$_2$ has increased in the presence of WS$_2$ due to ET. Thus, the emission intensity is larger for MoSe$_2$ in the heterostructure. On the other hand, emission from the WS$_2$ layer is partially quenched.

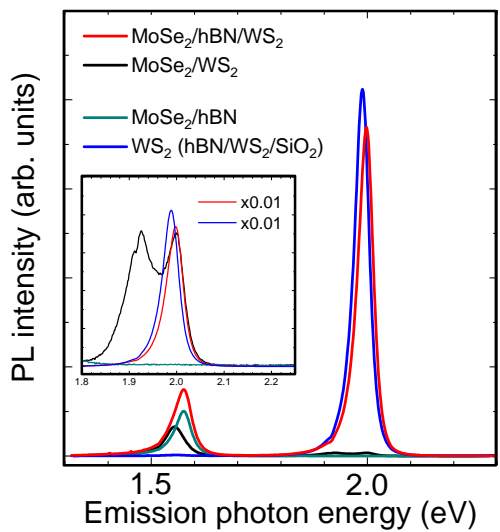

**Figure S11. PL spectra** for MoSe$_2$, MoSe$_2$/hBN/WS$_2$ heterobilayer, and MoSe$_2$/hBN/WS$_2$ heterotrilayer excited at 2.33 eV at room temperature, where the emission intensity is larger for MoSe$_2$. Emission from the WS$_2$ layer is partially quenched. The WS$_2$ layer is partially quenched.

**Supporting Note 11.    Comparison of energy transfer rate in various systems**

Van der Waals heterobilayer is an ideal system to investigate the ultimate limit of dipole−dipole energy transfer rates between two distinct semiconducting materials. This is because no two materials can be any closer than a van der Waals distance before electrons are shared between them in a form of covalent bonds. Determination of dipole−dipole energy transfer rate in this system is therefore important in its own right. We show that indeed the estimated energy transfer rate is among one of the fastest compared to those reported for other systems as shown in Figure S12.

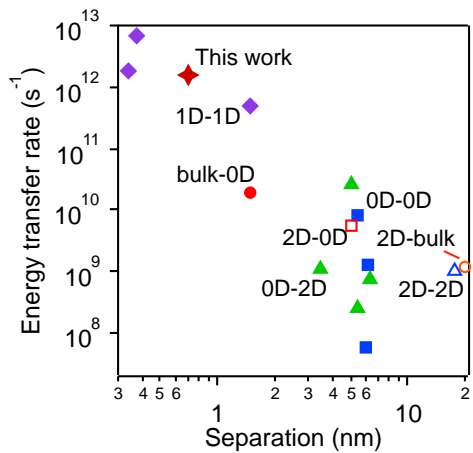

**Figure S12. Plots of energy transfer rate in various systems.** Quantum dots assemblies[43–45] (0D−0D), quantum dots−monolayer $MoS_2$[25, 46, 47], molecules−graphene[48] (0D−2D), carbon nanotubes[49-51] (1D−1D), quantum wells−quantum dots[52] (2D−0D), quantum wells[53] (2D−2D), quantum wells−organics[54] (2D−bulk), organics−quantum dots[55] (bulk−0D), and TMDs, this work, measured at 78 K.

**Table S1 | Typical Förster energy transfer rate in various hetero−systems.**

| Dimension | System | Separation (nm) | ET rate ($s^{-1}$) | Ref. |
|---|---|---|---|---|
| 0D−0D | QD assembly | 5.4 | $8.33 \times 10^9$ | 43 |
| 0D−0D | Close packed QD | 6 | $6.00 \times 10^7$ | 44 |
| 0D−0D | QD assembly | 6.2 | $1.33 \times 10^9$ | 45 |
| 0D−2D | QD−1L $MoS_2$ | 3.5 | $1.10 \times 10^9$ | 25 |
| 0D−2D | QD−1L $MoS_2$ | 6.3 | $7.70 \times 10^8$ | 46 |
| 0D−2D | QD−1L $MoS_2$ | 5.4 | $2.5 \times 10^8$ | 47 |
| 0D−2D | Molecules−graphene | 5 | $2.56 \times 10^{10}$ | 48 |
| 1D−1D | SWNT−SWNT | 1.5 | $5.00 \times 10^{11}$ | 51 |
| 1D−1D | DWNT | 0.38 | $6.67 \times 10^{12}$ | 50 |
| 1D−1D | SWNT bundles | 0.34 | $1.80 \times 10^{12}$ | 49 |
| 2D−0D | QW−QD | 5.1 | $5.80 \times 10^9$ | 52 |
| 2D−2D | QW−QW | 17.5 | $1.00 \times 10^9$ | 53 |
| 2D−bulk | QW−organic dye | 20 | $1.17 \times 10^9$ | 54 |
| Bulk−0D | Polymer−QD composite | 1.5 | $1.89 \times 10^{10}$ | 55 |